# Properties of Umbral Dots as Measured from the New Solar Telescope Data and MHD Simulations

A. Kilcik[1], V.B. Yurchyshyn[1], M. Rempel[2], V. Abramenko[1], R. Kitai[3], P.R. Goode[1,4], W. Cao[1,4], and H. Watanabe[3]

[1] Big Bear Solar Observatory, Big Bear City, CA 92314 USA
[2] High Altitude Observatory, NCAR, Boulder, CO 80307-3000, USA
[3] Kwasan and Hida Observatories, Kyoto University, Kyoto 607-8417, Japan
[4] New Jersey Institute of Technology, Center for Solar Research, Newark, NJ, USA

We studied bright umbral dots (UDs) detected in a moderate size sunspot and compared their statistical properties to recent MHD models. The study is based on high resolution data recorded by the New Solar Telescope at the Big Bear Solar Observatory and 3D MHD simulations of sunspots. Observed UDs, living longer than 150 s, were detected and tracked in a 46 min long data set, using an automatic detection code. Total 1553 (620) UDs were detected in the photospheric (low chromospheric) data. Our main findings are: i) none of the analyzed UDs is precisely circular, ii) the diameter-intensity relationship only holds in bright umbral areas, and iii) UD velocities are inversely related to their lifetime. While nearly all photospheric UDs can be identified in the low chromospheric images, some small closely spaced UDs appear in the low chromosphere as a single cluster. Slow moving and long living UDs seem to exist in both the low chromosphere and photosphere, while fast moving and short living UDs are mainly detected in the photospheric images. Comparison to the 3D MHD simulations showed that both types of UDs display, on average, very similar statistical characteristics. However, i) the average number of observed UDs per unit area is smaller than that of the model UDs, and ii) on average, the diameter of model UDs is slightly larger than that of observed ones.

**Introduction**

The umbra of sunspots is far from being homogeneous and it displays dynamic variations during its lifetime (see, e.g., a review by Solanki 2003). These inhomogeneities are often observed as bright, and mostly circular small patches embedded in the dark background (umbra). The patches were first described and named as umbral dots (UDs) by Danielson (1964). UDs appear in sunspots of all sizes, shapes, and intensities. The magnetic field in UDs is weaker than in the darker surroundings, and UDs show up-flows of a few hundred m s$^{-1}$ (Wiehr & Degenhardt 1993, Socas-Navarro et al. 2004, Rimmele 2004, 2008, Bharti 2007). According to realistic 3D simulations by Schüssler & Vögler (2006), UDs are due to a dominating convective energy being transported by narrow up-flow plumes with adjacent down-flows, which become almost field-free near the surface layer (see also Schlichenmaier 2009). As a result, bright UDs are observed in the dark central parts of the umbra. Revealing the nature of UDs is important for our comprehensive understanding of the mechanism of energy transport that defines the structure of a sunspot as we know it.

According to their origin, UDs are generally separated into two types: 1) central or umbral origin UDs and 2) peripheral or penumbral origin UDs (see Grossmann-Doerth et al. 1986, Kitai et al. 2007, Watanabe et al. 2010, and references therein). Peripheral

UDs are generally brighter than central ones (Sobotka & Hanslmeier 2005, Kitai et al. 2007). While the central UDs are mostly static, peripheral UDs move preferentially toward the center of the umbra with speeds less than 1.0 km s$^{-1}$ (Sobotka et al. 1997). The size, lifetime and velocity of UDs were extensively studied by many authors (e.g., Kitai et al. 2007, Riethmüller et al. 2008, Sobotka & Jurcak 2009). For instance, Riethmüller et al. (2008) and Rimmele (2008) found consistent values for UD sizes to be 0".2 – 0".5 and lifetimes of about 15 min. There are, however, some contradictory reports concerning the main statistical properties of UDs, since their determination depends on data quality, type of UDs and the detection and tracking methods used. Thus, Sobotka & Puschmann (2009) found a 4.5 min average lifetime, while Kitai et al. (2007) reported an average lifetime of 16 min. Similarly, the reported average diameter of UDs varies between 0".15 (~109 km) and 0".5 (~363 km) (Sobotka & Hanslmeier 2005, Kitai et al. 2007, Riethmüller et al. 2008, Sobotka & Puschmann 2009). Riethmüller et al. (2008) considered 2899 long living UDs (life time exceeding 150 s) and found their mean diameter to be 0".3 (229 km) with a mean velocity of 420 m s$^{-1}$.

In this study, we followed the approach of Riethmüller et al. (2008) and analyzed the lifetime, velocity, eccentricity, and diameter only of those UDs with lifetimes longer than 150 s. All parameters but eccentricity have been previously studied using observed (Kitai et al. 2007, Riethmüller et al. 2008, and references therein) and model (Schüssler & Vögler 2006) data. Schüssler & Vögler (2006) found that model UDs are not symmetrical but they rather display an elliptical shape. Here we analyze the shape of observed UDs for the first time using the New Solar Telescope and compare the results with recent numerical sunspot models (Rempel et al. 2009a,b, Rempel 2011).

**Observations and Data Reduction**

High resolution data for AR NOAA 11108, located at 20E30S heliographic coordinates, were obtained by the New Solar Telescope (NST, Goode et al. 2010, Cao et al. 2010) at the Big Bear Solar Observatory (BBSO). The 15 sec cadence photospheric data were recorded under good seeing conditions between 17:05 and 17:51 UT on September 20, 2010. The optical setup included a 1.6 m mirror and an adaptive optics system (AO) with a 97 actuators deformable mirror. The TiO data set, spanning 46 min time interval, consisted of total 184 AO corrected and speckle reconstructed images (Wöger & von der Lühe 2007) obtained with a 0.3 nm passband TiO filter centered at the 705.7 nm spectral line. This spectral line is sensitive to temperature, and it is exceptionally suitable for observing sunspot umbra and penumbra (Berdyugina et al. 2003, Riethmüller et al. 2008, Abramenko et al. 2010). The field of view (FOV) of the TiO imager was 77" x 77" and the pixel scale of the PCO.2000 camera we used was 0".0375. This pixel size is 2.9 times smaller than the Rayleigh diffraction limit of the NST ($\theta_1 = 1.22\lambda/D = 0".11 = 77$ km) and 2.5 times smaller than the FWHM of the smallest resolved feature ($\theta_2 = 1.03\lambda/D = 65$ km, Abramenko et al. 2010).

Along with the photospheric data, Hα images were simultaneously obtained with the Narrowband Filter Imager (NFI). Due to the time needed to tune the 0.025 nm bandpass Lyot filter between several wavelengths, the Hα data set consisted of many groups of 4 images, taken with an 8 s cadence, intermittent with 20 s gaps. Here we will use only images acquired at Hα -0.075 nm.

Leenararts et al. (2006) calculated the line formation height of Hα based on quiet-sun atmospheric model and concluded that the wing of Hα contains significant contribution from the photosphere, so that granulation is apparent in the far blue wing Hα images. Vernazza, Avrett, & Loeser (1981) reported that Hα wings form in the 500km layer above the photosphere. In this paper, for simplicity and clarity sake we will refer to the UD seen in the off-band Hα images as "low chromospheric" UD.

The data sets were further processed as follows: i) all images were aligned and de-stretched to remove the residual image distortion due to seeing and image jittering, ii) intensity of each image was adjusted to the average level of the set, iii) to enhance umbral structure a standard masking procedure was applied by using a high degree smoothing, and iv) low pass fast Fourier and band pass filters were applied to reduce the noise level in the masked images. The band pass filter utilizes a wavelet technique based on the convolution with a ''Mexican hat'' kernel to remove random digitization and the background noise (Crocker and Grier, 1996). The resulting data cubes have a FOV of 675 pixels x 675 pixels (25".3 x 25".3) and include the entire umbrae and a part of the penumbra (Figures 1 and 2). Figure 1 shows Fourier filtered images, while Figure 2 shows the result of masking, band passing and UD detection procedures.

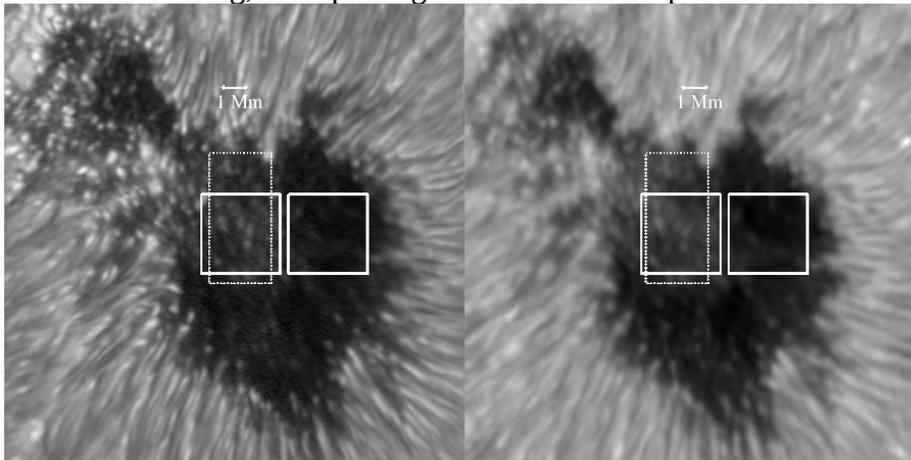

*Figure 1. Analyzed co-temporal photospheric (left) and low chromospheric (right) images. Small boxes describe selected brighter and dimmer areas used for analysis, while the dotted rectangle indicates a part of the umbra plotted in Figure 4.*

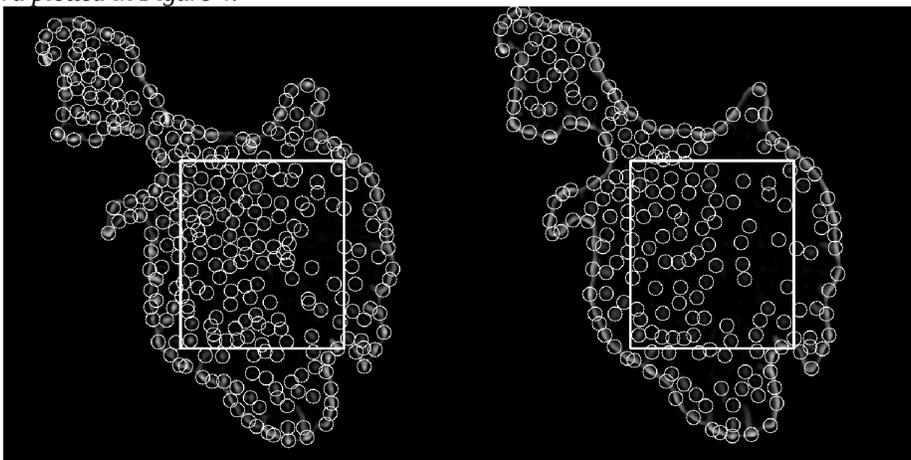

*Figure 2. Standard masked and bandpass filtered photospheric (left) and low chromospheric (right) images. The circles indicate positions of the detected UDs (but not their size). The rectangles show the selected small area used in the Results section.*

**Method and Analysis**

To detect and track UDs we used an IDL code by Crocker & Hoffman (2007), originally written for blood cell detection and tracking. The detection method is based on measurements of local intensity of analyzed data. We began the analysis by choosing an initial value of the local area diameter, which was set to 17 pixels (0".64) for both photospheric and low chromospheric data. The local area diameter should be large enough to encompass the largest UD (0".43) in the data set. In addition, the total intensity of the local area is also used as a criterion for UD detection. To remove the noise in the data, the total intensity threshold is set by visually inspecting the best images in the data sets. The best local area minimum intensity threshold was set to 10000 DN for both photospheric and low chromospheric images. Thus if the total intensity of the local area is less than 10000 DN this local area will not considered as a UD. For comparison, the average intensities of the investigated local areas were 15842 DN and 16276 DN for the photospheric and low chromospheric data, respectively. The shortcoming of the thresholding approach is that the dimmest UDs were excluded from the analysis (full effect of the thresholding will be discussed further in the text).

The outcome of the UD detection routine is shown in Figure 2, where the equal radius circles indicate the position of the detected UD (but not their size). UD tracking procedure is based on the area overlap approach. To track the UDs, we applied the following criteria: i) coordinates of the center of a UD in the current image should be within a circle of radius *r*, equal to the radius of a given UD, centered at the central coordinates of a given UD in the next image; i) each UD should be present in 10 subsequent images, i.e., only UDs with life time > 150 s were tracked, and ii) if an UD disappears and then appears again at the same location within the following 5 images (corresponds to 75 s interval), it is considered to be the same entity. To illustrate the performance of the detection routine, in Figure 3 we over-plot a small fragment of the umbra with ellipses, whose central coordinates, size and orientation were determined from the detection code. A fragment of the umbra over-plotted with ellipses, which are the best fit to the observed UDs. Note that the ellipses encircle the UDs at the intensity level much lower than those represented with the gray-scale intensities, which explains the difference between the size of UDs and ellipses.

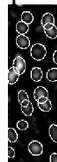

*Figure 3. A fragment of the umbra over-plotted with ellipses, which are the best fit to the observed UDs.*

The diameter and eccentricity of a UD were derived from the best fit of an ellipse to the selected intensity element, with the diameter being measured as the major semi-axis of an ellipse. We also calculated UD's minor axis diameter from the ellipse eccentricity and major semi-axis (see Figure 5, 11, and 13). The linear velocity was calculated by measuring the total displacement of a UD over its lifetime. Here we used average values of all these parameters over lifetimes.

To compare statistical and physical properties of photospheric and low chromospheric UDs, a small rectangular area (240 pixels x 276 pixels) in the center of the umbra was selected (see Fig 2). Finally, we separated all photospheric UDs belonging to this small area into two classes. The first class included the UDs that only appeared in the photosphere, while the second class included events that existed in both photospheric and low chromospheric images. To identify the same UDs in the photosphere and chromosphere, we used their average central coordinates ($x_c$ and $y_c$) and the appearance/disappearance times, determined at both atmospheric layers. We accepted that the difference between the low chromospheric and photospheric coordinates should not exceed the selected local area diameter of the photospheric UDs (17 pixels), while the difference between their appearance and disappearance times should not exceed 4 min.

**Results**

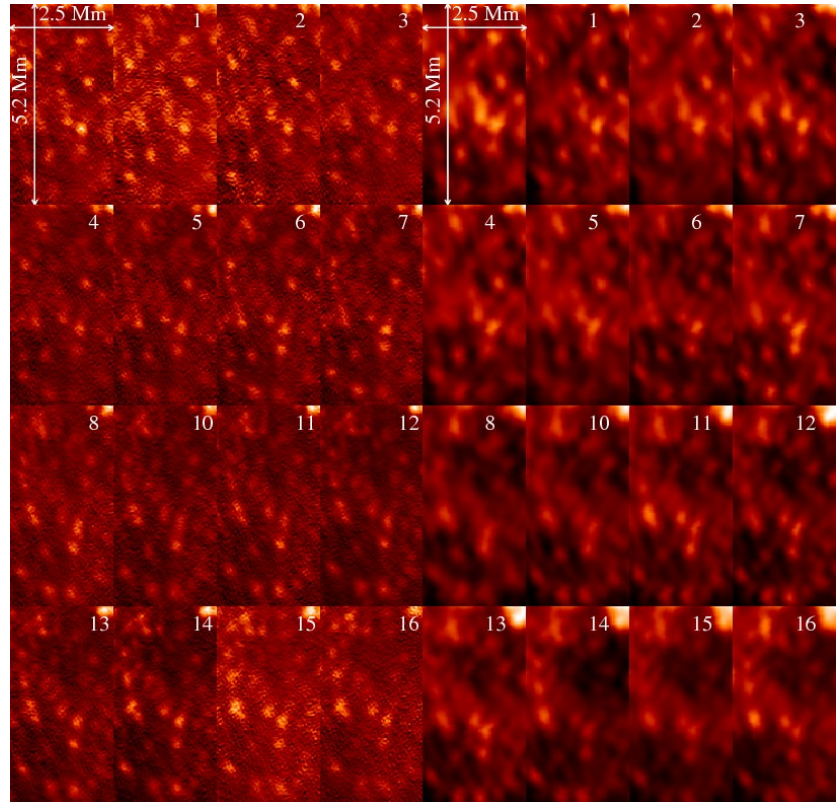

*Figure 4. Evolution of photospheric (left) and low chromospheric (right) UDs spanning approximately 17 min of time. The time interval between each frame is approximately 1 min and the starting time is 17:28:30 UT. The number in the upper right corner of each frame indicates the time in minutes from the starting time.*

In Figure 4, we present approximately a 17 minutes time evolution of UDs observed at the photospheric 705.7 nm spectral line and in the blue wing of the Halpha line. Each small panel represents the area outlined by the dotted rectangle in Fig 1. The time step between images is about 1 minute. Most of the UDs can be identified in both images with the difference that the photospheric UDs are in general better resolved as compared to the low chromospheric ones. Sometimes, it appears that several photospheric UDs shared one entity of low chromospheric UDs. Both the lower spatial resolution of the chromospheric data as well as the real difference in physical parameters may contribute to the fact that low chromospheric UDs appear to be less well-defined. The UD's filling factor is not equally distributed over the umbra, either. While the upper part of the sub-image in Figure 4 is tightly packed with bright UDs, the middle section of the frame is either nearly devoid of them or the surface density of UDs is much lower. This non-uniformity is thought to be reflecting the non-uniformity of the magnetic field density inside the umbra.

Table 1. The number of detected UDs and their average parameters listed separately for all photospheric (APhUD) and low chromospheric (ALChUD) UDs, as well as for the events measured inside the selected area (SPhUD and SLChUD).

|  | Number of UDs | Diameter, arcsec | Eccentricity | Lifetime, min | Linear velocity km s$^{-1}$ |
|---|---|---|---|---|---|
| APhUD | 1553 | 0.35 | 0.74 | 8.19 | 0.45 |
| SPhUD | 530 | 0.35 | 0.74 | 7.99 | 0.43 |
| ALChUD | 620 | 0.36 | 0.75 | 10.42 | 0.34 |
| SLChUD | 247 | 0.35 | 0.74 | 10.23 | 0.32 |

In Table 1, we list statistical properties of UDs such as the diameter, eccentricity, life time and linear velocity determined from photospheric and low chromospheric data and in Figure 5, we plot probability distribution functions (PDFs) for these parameters. Note that the eccentricity of an UD describes its shape: a unit eccentricity indicates a circular UD, while zero eccentricity indicates a line.

As follows from the Table 1, all detected UDs show an average diameter of 0".35, a lifetime of about 9 min and a velocity of about 0.39 km s$^{-1}$. While the lifetime of low chromospheric UDs is, on average, slightly bigger than photospheric ones, their average linear velocity is smaller. The average diameter and eccentricity of UDs do not seem to show any atmospheric layer dependency. Finally, the number of detected UDs is much larger in the photosphere than in the low chromosphere. In Figure 5, we plot the corresponding probability distribution functions (PDFs) for photospheric (solid) and low chromospheric (dotted) data. The upper left panel shows the diameter PDFs, which is slightly asymmetrical and has a more pronounced tail at smaller scales. Both photospheric and low chromospheric PDFs have very similar characteristics. The gray lines show PDFs of UD's diameter, measured as the minor semi-axis of an ellipse fitted to a UD. While the general shape of the minor diameter PDFs resembles that of the major diameter PDFs, the most prominent difference is that the minor diameter PDFs are centered, as expected, at smaller spatial scales of 0".26. Therefore, we may conclude that depending on the method of UD size measurements their average size may range between 0".26 and 0".39. The upper right panels plots eccentricity of UDs. Essentially, all UDs are elongated and none of them was found to have a circular shape. A small fraction of UDs appear to be well-elongated and they most probably belong to the peripheral type of events, which are often observed near the umbra-penumbrae boundary. The two lower panels plot PDFs for the lifetime and velocity of UDs. They are not symmetrical, either, and show a somewhat heavy tail on large-scales, which is reminiscent of a log-normal distribution. Again, both photospheric and low chromospheric PDFs display similar characteristics, although low chromospheric UDs tend, on average, to live longer and move slower.

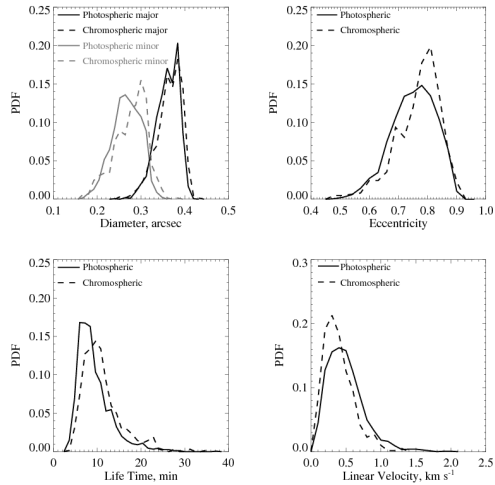

*Figure 5. PDFs for diameter (major axes of ellipse), eccentricity, lifetime and linear velocity of all photospheric (solid line) and low chromospheric (dashed line) UDs. The gray lines in the diameter PDF (upper left panel) show the diameters obtained from the minor axis of an ellipse.*

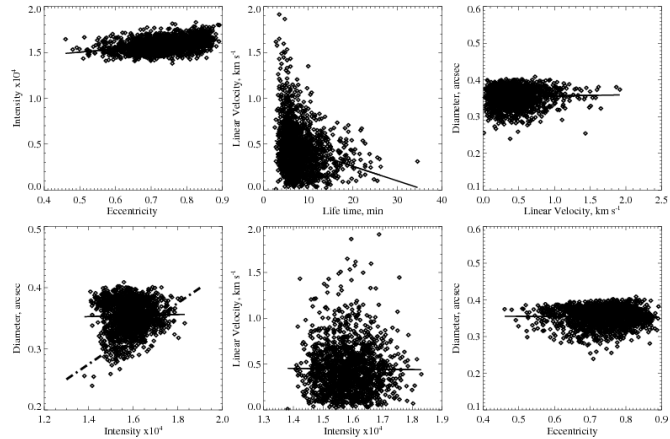

*Figure 6. Relationship between intensity, diameter, eccentricity, lifetime and velocity of photospheric UDs. The solid line is the best linear fit to the data points. Long dashed–dotted line in the lower left panel indicates data points corresponding to the UD population associated with bright areas inside the sunspot's umbra.*

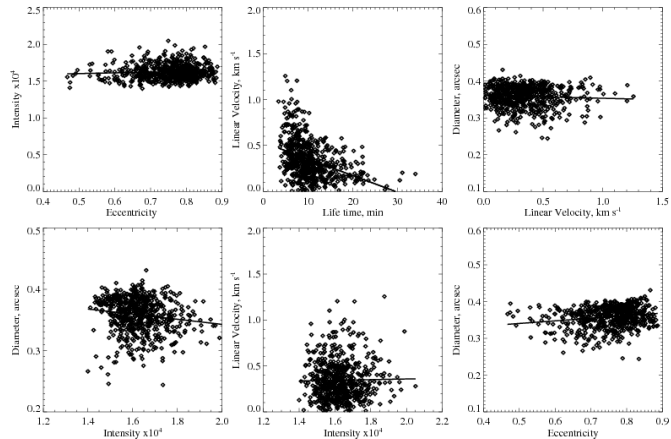

*Figure 7. Relationship between intensity, diameter, eccentricity, lifetime and velocity of low chromospheric UDs. The solid line is the best linear fit to the data points.*

In Figures 6 and 7, we plot the inferred parameters against each other in order to address the possibility of any connection between them. These plots show that fast moving photospheric and low chromospheric UDs have, on average, shorter lifetimes (upper middle panel) and that the intensity and diameter of UDs show only a weak connection to their eccentricity. The correlation coefficients of all compared parameters are given in Table 2. The relationship between the intensity and the diameter of UDs (lower left panel) appears non-trivial. In this panel, the nearly horizontal line is the best fit to all data points, which appears to consist of two different subsets. For one subset, there appears to be no relationship between the diameter and intensity, while the other subset seems to have a well-pronounced, direct relationship. Dash-dotted line in the lower left panel of Figure 6 indicates this second subset of data points. This line is only to guide the eye and it was produced without using any numerical fitting routines.

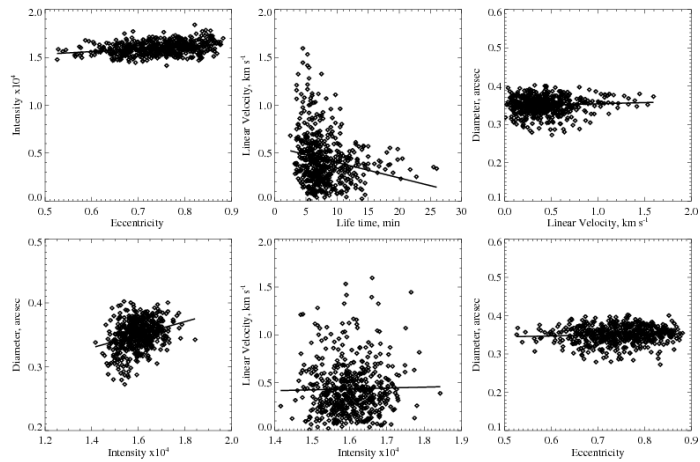

*Figure 8. Relationship between intensity, diameter, eccentricity, lifetime and velocity determined for the selected photospheric UDs (total 530 UDs) detected inside the rectangular area shown in Figure 2. The solid line shows the best linear fit.*

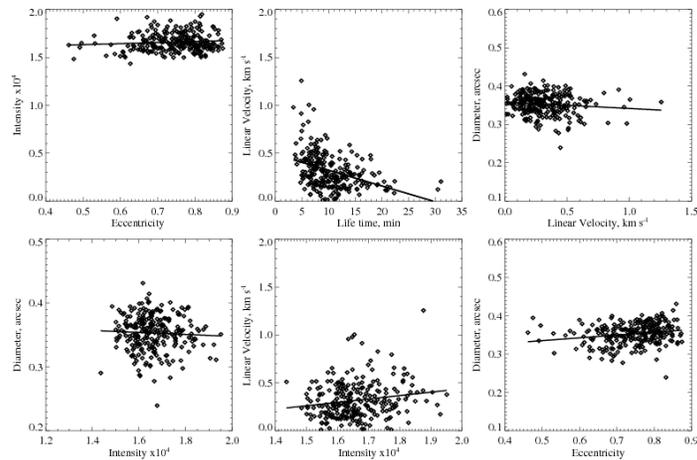

*Figure 9. Relationship between intensity, diameter, eccentricity, lifetime and velocity determined for the selected low chromospheric UDs (total 247 UDs) detected inside the rectangular area shown in Figure 2. The solid line shows the best linear fit.*

*Table 2: Correlation coefficients between various parameters of UDs. Abbreviation is the same as in Table I.*

|  | Intensity-Diameter | Intensity-Velocity | Diameter-Eccentricity | Eccentricity-Intensity | Lifetime-Velocity | Velocity-Diameter |
|---|---|---|---|---|---|---|
| APhUD | 0.02 | -0.01 | - 0.01 | 0.35 | -0.23 | 0.03 |
| ALChUD | -0.15 | 0.02 | 0.19 | 0.07 | -0.38 | -0.06 |
| SPhUD | 0.30 | 0.02 | 0.07 | 0.29 | -0.21 | 0.07 |
| SLChUD | -0.05 | 0.17 | 0.21 | 0.09 | -0.38 | -0.11 |
| Model UD | 0.57 | -0.01 | -0.17 | -0.02 | -0.35 | 0.20 |

In Figures 8 and 9, we plot the same parameters, but for selected UDs, i.e., UDs located within the area encompassed by the white box in Figure 2. In general, all the parameters for this subset show relationships similar to those determined for all UDs with some differences (see Table 2). The selected photospheric UDs now do show a well-pronounced direct relationship between their intensity and diameter (Figure 8, lower left panel). The same inverse relationship for the selected low chromospheric UDs was found. This change of properties of photospheric UDs and the diameter-intensity plot in Figure 6 (lower left panel) seem to suggest that there could be two different populations of UDs present in the photospheric umbra. To explore this divergence of results, we defined two squared areas (4".35 x 4".35) inside the umbra (solid boxes in Figure 1). One box encompassed the brightest part of the umbra, while the other included the dimmest part of the umbra. We then re-plotted the above mentioned relationships separately for the bright and dim areas (see Figure 10), and found that the UDs belonging to the brighter area show a direct relationship with their diameter, while the dimmer area UDs do not show this relationship at all. We thus speculate that the intensity–size relationship is apparently only applicable inside the brighter parts of the sunspot's umbra (see Table 2) where the magnetic field intensity is lower when compared to the field intensity in the dimmest parts.

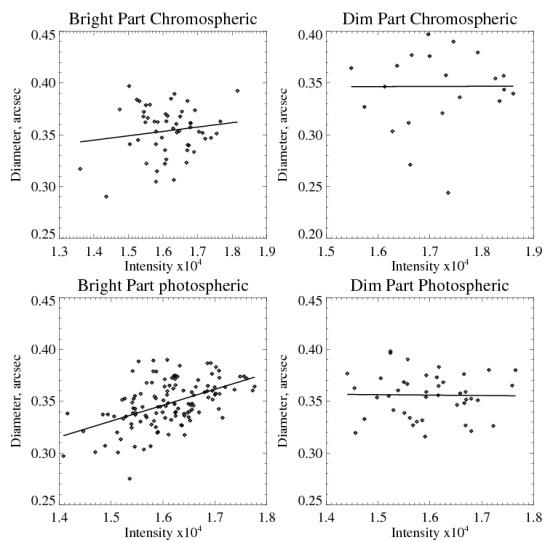

*Figure 10. Relationship between the intensity and diameter of UDs found in the brightest and dimmest parts of the umbra. The solid line shows the best linear fit. The correlation coefficients between the diameter and intensity of UDs are 0.51, -0.02, 0.15, and 0.0 for the brightest and dimmest photospheric UDs and brightest and dimmest low chromospheric UDs, respectively.*

Finally, we separated the selected photospheric UDs, detected inside the box (Figure 2), into two subsets: one subset included those UDs that only existed in the photosphere, while the second subset contained UDs that appeared in both photospheric and low chromospheric data. In Figure 11, we separately plot PDFs of their parameters. We find that about 85% of low chromospheric UDs had a counterpart in the photosphere, while only 38% of photospheric UDs appear in the low chromosphere. This apparent disconnect between photospheric and low chromospheric data may be at least partially explained by the fact that many low chromospheric UDs appear to have a weaker contrast. This means that two neighboring UDs, well separated in the photospheric images, appear in the low chromosphere as one large bright entity, therefore it cannot be identified as an UD since its size exceeds the maximum local area diameter. We thus conclude that a larger number of detected photospheric UDs (as compared to the low chromosphere) is most probably caused by this effect. Moreover, visual inspection of Figure 4 also shows that nearly all photospheric UDs can be identified in the chromospheric images.

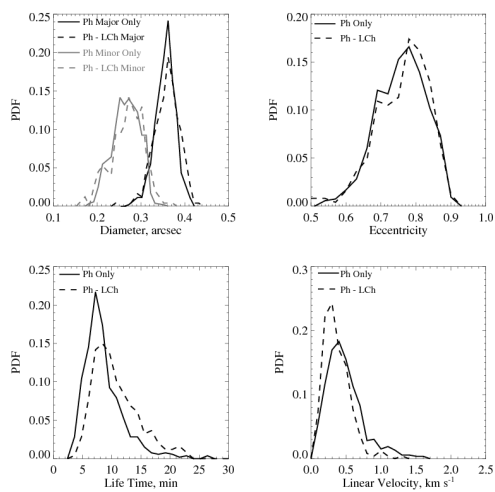

*Figure 11. PDFs for diameter, eccentricity, lifetime and linear velocity determined for two subsets of selected UDs. Solid lines show PDFs for photospheric UDs only (Ph only), while the dashed line represents PDFs for the photospheric UDs, which have low chromospheric counterparts (Ph – LCh). The light lines in the diameter PDF show the minor axis diameter of the ellipse.*

The lower left panel in Figure 11 indicates that the fraction of short living (4-8 min) UDs in the low chromosphere is much smaller than that in the photosphere, while long living events are nearly the same. The velocity PDFs indicate that the chromospheric UDs are mostly slow moving events, while the diameter and eccentricity PDFs are similar.

**Comparison with Model Data**
Here we compare statistical parameters of observed and simulated UDs. The sunspot model we analyze has been computed using the approach described in Rempel at al. (2009a, b) with a slightly different setup, which will be described in detail in a forthcoming paper. The domain size is 49.152x49.152x6.144 $Mm^3$, and the numerical grid resolution 32x32x16 $km^3$. The net flux within the domain is $1.2 \, 10^{22}$ Mx, of which about $10^{22}$ Mx a sunspot with a radius of about 18 Mm (umbra radius of about 8 Mm).

The field strength at the center of the spot is about 3.3 kG. The simulation was initially run at a lower resolution of 48x48x24 km³ for 3.3 hours, followed by about 2.2 hours at 32x32x16 km³, which were analyzed here. This simulation was followed by 1 hour in 16x16x12 km³ resolution, an intensity snapshot from that sequence can be found in Rempel (2011). For our analysis we selected a 51 min long model data series run with a spatial resolution 32x32x16 km³.

In Figure 12, we show a snapshot of a model umbra and the corresponding detected Uds.

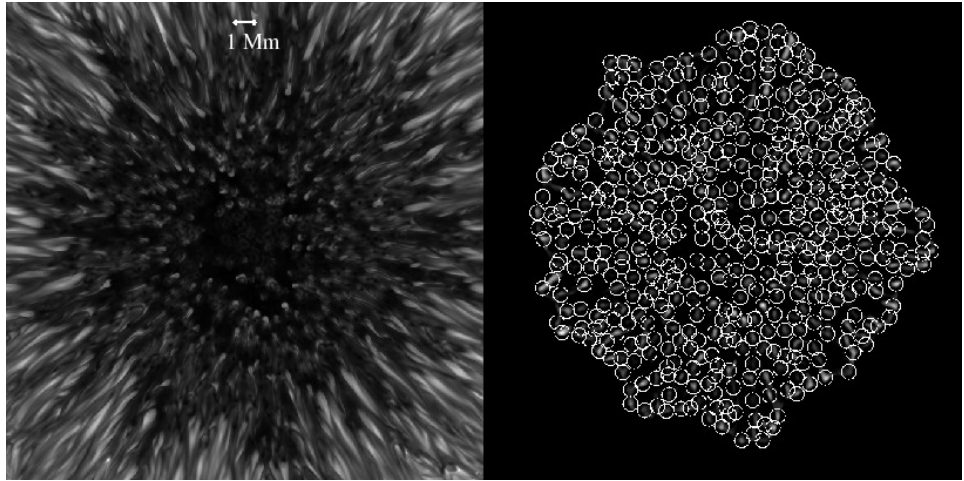

*Figure12. Analyzed model umbra (left) and the corresponding detected UDs (right). The field of view of the image is 29".6 x 29".6.*

Visual inspection and comparison of Figures 1 and 12 shows that i) the model umbra appears to have a higher surface density of UDs; ii) model UDs are not circular and appear to be elongated along the sunspot's radius; iii) model UDs show a finer structure not detected in the utilized observed data; and iv) similarly to observations, the model umbra contains zones of low and high surface density of UDs.

*Table 3. Average parameters of observed and modeled UDs.*

|  | Number of UDs | UD surface density, Mm$^{-2}$ | Diameter, arcsec | Eccentricity | Lifetime, min | Linear Velocity, km s$^{-1}$ |
|---|---|---|---|---|---|---|
| Observations | 1553 | 1.9 | 0.35 | 0.74 | 8.19 | 0.45 |
| Model All Umbra | 3232 | 2.4 | 0.41 | 0.72 | 12.9 | 0.35 |
| Model Central Umbra | 513 | 2.3 | 0.37 | 0.76 | 16.2 | 0.20 |

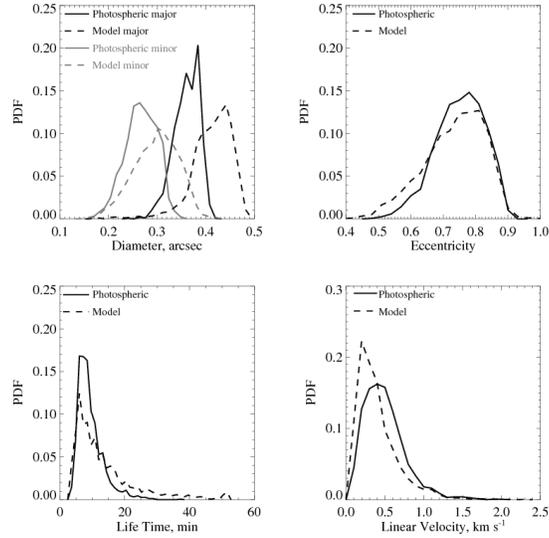

*Figure 13. PDFs for model (dashed line) and observed photospheric (solid line) UDs. The gray lines in the upper left panel show minor diameter PDFs measured as the length of the minor axis of an ellipse.*

In Table 3 and Figure 13, we list average parameters of model UDs and plot corresponding PDFs. For comparison, we also list the same parameters derived from observed UDs and as model UDs located near the center of the umbra. In general, the statistics of model and observed UDs are very similar, with model UDs having slightly larger diameters and lifetime, while the velocities are lower. The average diameter and eccentricity of the central model UDs are closer to the observed parameters, while their life time is about twice as long and their linear velocity is half of the observed ones.

To calculate the UD surface density, we used one high quality observed image and one randomly selected model image. The model UD surface density was found, on average, to be slightly higher (2.4 $Mm^{-2}$) than the observed UD density (1.9 $Mm^{-2}$). The entire model umbra and the central part of the umbra both have nearly same density (2.4 $Mm^{-2}$ and 2.3 $Mm^{-2}$, correspondingly). The surface density of observed UDs inside the small box encompassing the brightest part of the umbra (Figure 1) is larger (2.8 UD per $Mm^{-2}$) than the model density, however, the UD surface density found within the darkest part of the umbra was only 1.3 UD per $Mm^{-2}$.

The profiles of model and observed PDFs have similar shapes, indicating that i) NST data allowed us to detect and measure most of the photospheric UDs, but the smallest ones (model PDF extends to smaller scales, upper left panel), and ii) the sunspot model seems to properly reproduce not only the general appearance of a sunspot, but also captures all essential and relevant physical processes on the scale of umbral fine structure.

Another noticeable distinction is that the model velocity PDF peaked at smaller velocity intervals. It appears that the model UDs is dominated by low velocity and stationary UDs. In the previous section, based on observations, we argued that slow moving UDs usually live longer. This inference is also supported by the lifetime PDF determined from the model data: it is apparent that the number of model UDs living longer than 10 min significantly exceeds that of the observed UDs.

In Figure 14, we plot modeled UD parameters against each other. In general, all model relationships are very similar to those determined for photospheric UDs (see Figure 6, and Table 3). Lifetime and velocity of model UDs show an inverse relationship, while their linear velocity is increasing with size (upper right). Model data also seem to indicate that the eccentricity of at least some UDs directly depends on their diameter.

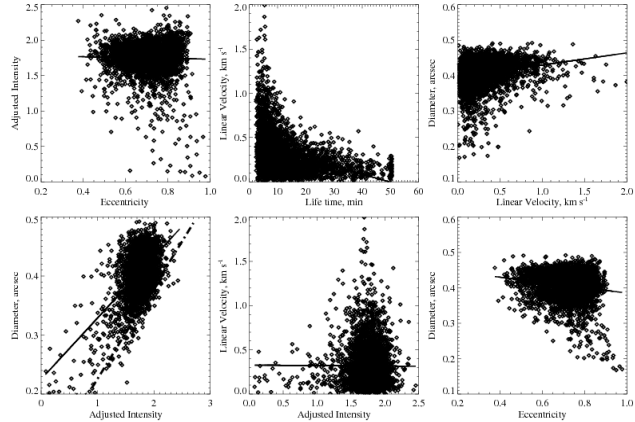

*Figure 14. Relationship between intensity, diameter, eccentricity, lifetime and velocity determined for the model sunspot UDs (total 3223 UDs). The solid line is the best linear fit to the data points. Long dashed – dotted line in the lower left panel is the same as in the lower left panel of Figure 6 and it indicates data points corresponding to the UD population associated with the bright parts of the observed umbra.*

Similarly to the observed data, the diameter of model UDs is directly related to their intensities (lower left panel). However, there are some differences. In the previous section, based on observed data, we speculated that there could be two distinct populations of UDs: i) UDs that comprise the brightest parts of the sunspot umbra and show intensity–size relationship (see dashed line in Figure 6, lower left panel) and ii) UDs that mainly populate darkest parts of the umbra and their intensity does seem to be related to their diameter. In Figure 14 (lower left), the dash-dotted line is the same as in Figure 6, while the solid line is the best linear fit to the data points. As one can see, the dash-dotted lines produced from observed data, relatively well agrees with model data pionts. Also, model data points do not show the subset associated with darkest parts of the umbra. It thus appears that the model data is heavily dominated by only one type of UDs associated with bright umbra. This may not be surprising considering that the dark area in the model umbra occupies only a small fraction of the total area, so that the second type UDs may be underrepresented.

Finally, to explore dependence of UD parameters on the model resolution, we calculated the average size, lifetime, velocity and eccentricity by using the higher, 16x16x12 km$^3$, resolution model run. The only difference between two data sets is the 16 km-resolution model produced about 18 % more UDs than the 32 km-resolution model. All the other average parameters were almost the same.

**Conclusion and Discussion**
In this study, we analyzed properties of UDs detected in the umbra of the main spot of AR NOAA 11108 observed by the NST on September 20, 2010. Our analysis was complemented by 3D MHD model data that included a realistic equation of state and radiative transfer. Our new findings are as follows:

1. None of the analyzed UDs is of an exact circular shape. The average photospheric and low chromospheric eccentricity is 0.74 and 0.75, respectively, with 0.08 standard deviation. This is the first observational confirmation of the earlier theoretical findings.
2. We found that the diameter-intensity relationship appears to work for only brighter umbral areas, where the diameter increases with UD's intensity.
3. We further confirm that the diameter of photospheric UDs varies from 0".23 to 0".41.
4. UD velocities show clear anti-correlation with the lifetime: fast moving events have shorter lives.
5. While photospheric UDs are dominated by the population of fast moving and short lived UDs, the low chromospheric UDs are mostly slow moving and long lived structures.
6. Not all photospheric UDs protrude high enough to reach the low chromosphe. For about 85% of low chromospheric UDs we were able to detect their counterparts in the photosphere, while for only 38% of photospheric UDs counterparts were detected in the low chromosphere.
7. PDFs for model UD parameters show strong similarity with the observed PDFs.
8. The average diameter of model UDs (for the model with 32x32x16 $km^3$) was found to be slightly larger than that found from observations.
9. The average number of observed UDs per unit area (surface density) determined over the entire observed umbra is slightly smaller than that of the model UDs. However, the brightest part of the observed umbra was found to have surface density about 17 % higher than the model ones, while the darkest observed umbra have a UD surface density 46 % lower that the model umbra.

The NST data allowed us to address the shape of the UDs for the first time. We found that UD's eccentricity varies between 0.56 and 0.89 with 0.75 and 0.74 average values for low chromospheric and photospheric UDs, respectively. The majority of UDs are oblate and none of the detected and tracked UDs had an exact circular shape ($\varepsilon = 1$). This agrees with Schüssler & Vögler (2006) simulations showing that UDs are not circular. We further find that the calculated eccentricities show a weak inverse relationship with the size and the intensity: bigger and brighter UDs tend be rounder.

We found a strong anti-correlation between UD lifetimes and their velocities. Muller (1973) reported an anti-correlation between the velocity and the lifetime for UDs of penumbral origin. Watanabe et al. (2010) analyzed a rapidly moving UD, whose lifetime was much shorter (8.7 min) than the generally reported value (~15 min), while its size was nearly the same as the average size of UDs (240 km). Our findings are in agreement with above reports. Moreover, Watanabe et al. (2009) analyzed Hinode sunspot data for AR NOAA 10944 and reported that in regions with strongly inclined field the speed of UDs increases with the associated magnetic field inclination angle, while the direction of their displacement is nearly parallel to the direction of the horizontal component of the magnetic field. We thus speculate that fast moving UDs are associated with a strong horizontal component of the magnetic field and/or strongly inclined fields.

Kitai (1986) compared chromospheric and photospheric UDs living longer than 10 min and reported that chromospheric UDs have characteristics similar to the photospheric ones, from the viewpoint of proper motion. The authors also suggested that chromospheric UDs, or at least some of them, are chromospheric counterparts of photospheric UDs. Our analysis supports Kitai's (1986) suggestion. Although the number of automatically detected photospheric UDs significantly exceeds the number of those in the low chromosphere, visual analysis of photospheric and low chromospheric umbrae indicates that nearly all photospheric UDs can be identified in the low chromospheric images. The disparity between the numbers of detected UDs arises from the fact closely spaced UDs appear in chromospheric images as an unresolved large cluster, so that the individual UDs within this cluster cannot be detected. This finding is in disagreement with Kitai (1986) who reported that chromospheric UDs are more numerous. One possible explanation of the discrepancy is that while these clusters of UDs can be seen in the lower resolution low chromospheric data, the corresponding small UDs were not present in the photospheric data.

The smallest UDs detected in our study have major diameter of 0"23 and the minor diameter of 0".15, which is above the NST diffraction limit (0".11). These lower bounds of observed UD size agree well with the measurements obtained for model data, which may be evidence that the NST has probably resolved the smallest UDs. On the other hand, we note that the automatic detection method we used here is based on setting an intensity threshold. Even though we used as low threshold as the data quality allows, there is a possibility that smaller UDs do exist, however their contrast and lifetime should then be below the detection limit of the data and methods. Also, the observed diameter PDF displays a weak asymmetry resulting in a heavier tail at smaller scales on the spatial range. The above facts lead us to suggest that there might be an undetected population of very small and low contrast UDs.

Kitai et al. (2007) analyzed three days of Hinode observations of AR NOAA10944 and reported that UDs lifetimes range from 4 to 40 min with an 14.8 min average. The size of UDs varies between 0".32 and 0".5, while their average velocity is 0.5 km s$^{-1}$. Riethmüller et al. (2008) analyzed 110 min time series obtained by the Swedish Solar Telescope. These authors analyzed UDs in two classes, which included all UDs and only those UDs that lived longer than 150 s. They reported 10.5 min average lifetime for the class of longer living UDs. Their average diameter and the average speed were found to be 0".375 (272 ± 53 km) and 0.42±0.02 km s$^{-1}$, respectively. Based on the 15 s cadence data, we report that the lifetime of photospheric UDs varies between 2.5 and 34.5 min, with 8.2 min average value, while the diameter varies between 0".23 and 0".41 with 0".35 (253.8 km) average. The velocity ranges from 0.02 km s$^{-1}$ to 1.92 km s$^{-1}$ with 0.45 km s$^{-1}$ average. In case of low chromospheric UDs the average lifetime and diameter is 10.42 min and 0".36 (bigger than photospheric values), respectively, while the linear velocity is smaller (0.34 km s$^{-1}$) as compared to the photospheric UDs. All averaged photospheric UD parameters that we report here are, in general, in a good agreement with those published earlier.

It was shown in earlier studies that the central UDs are less dynamic compared to the peripheral ones (Kitai 1986, Ewell 1992, Sobotka et al. 1997, Kitai et al. 2007, and references therein). Although we did not separate the detected event into two groups based on their distance from the center of the sunspot, our analysis of selected UDs (see

Table 1) shows that the earlier findings are true only in a statistical sense.

A positive relationship between the intensity and diameter of UDs had previously been reported by Tritschler & Schmidt (2002). Riethmüller et al. (2008) reported that the brighter UDs are on average a bit bigger than the small ones. Recently, Bharti et al. (2010) concluded from realistic 3D radiative MHD simulation that larger UDs tend to be brighter and live longer. We found that the intensity–diameter relationship holds only for UDs belonging to the brightest parts of the umbra. While our results confirm previous reports by Tritschler & Schmidt (2002), Riethmüller et al (2008) and Bharti et al (2010), on the relationship between UD intensity and diameter, we did not find any reliable relationship between the UD size and their life time.

A very good agreement between observed UD parameters and those determined from a 3D MHD model (Rempel et al. 2009a, b, Rempel 2011), validates the utilized UD detection method utilized, as well as illustrating that the model adequately represents sunspots. Bharti et al. (2010) performed an analysis of statistical properties of model UDs and found their average lifetime to be about 25-30 min, which is much longer that the lifetimes derived in this study. Moreover, the lifetime distribution function in Bharti et al. (2010) significantly differs from those in Figure 13 (lower left panel). One possible explanation of the disagreement may be due to the fact that Bharti et al. (2010) focused entirely on central UDs, where a vertical field was imposed: our estimates of the lifetime of central model UDs do show longer averaged life time (16 min), however, it is still far too short compared to Bharti et al. (2010). Additional differences exist in the position and treatment of the bottom boundary (here we have a closed boundary located about 4.5 Mm below the average $\tau = 1$ level in the umbra vs. an open boundary condition in about 1.2 Mm depth in Schüssler & Vögler (2006)), as well as the treatment of numerical diffusivities. We speculate that in particular, the differences in the bottom boundary could influence the lifetime of UDs. A determination of the exact cause would require additional model runs.

The remaining differences we found between observations and MHD simulations could have several causes. It is possible that some of these discrepancies are simply due to differences in the overall properties between our observed and simulated sunspot.

The size of umbral dots in the numerical models could be affected to some degree by the numerical resolution. Since about 5-10 grid points are needed to resolve a UD, this sets a numerical minimum size of about 0".2-0".4 for the simulation with 32 km resolution and 0".1-0".2 for the simulation with 16 km resolution. While these numbers are not too different from the sizes of UDs we found, we did not see a dramatic change between the two simulations. This indicates at least some robustness with regard to resolution, although a higher resolution than 16 km is certainly desirable for future numerical studies of UDs.

The difference in number density can be related to differences between the overall properties of the observed vs. the simulated sunspot. The model data produced UDs with nearly homogeneous number density over the entire umbra, while the observed sunspot displays highly non-homogeneous number density (see Fig 2 and Fig 12). Rempel et al. (2009a) found a strong dependence in the overall number of UDs on the field strength in the umbra by simulating a pair of sunspots with 3.2 and 4.2 kG central field strength. In addition, resolution effects (and closely related numerical diffusivities) could play a role. We also note that some differences can arise from image degradation, stray light and noise in the observations, which can hide some of the fainter UDs in observations. This can also lead to some differences in sizes and lifetimes. If fainter parts of UDs remain hidden, the observed size is smaller and the observed lifetime is shorter since forming UDs will become visible later, while decaying UDs disappear earlier. On the other hand, the convolution with the telescope point spread function can also broaden features. We postpone a detailed analysis of these effects to a future paper.

Authors thank J. Crocker of Pennsylvania Muscle Institute, and E. Weeks of Emory University for the IDL code and their assistance. This research was supported by NASA grants GI NNX08AJ20G, LWS NNX08AQ89G, and NNX08BA22G as well as NSF ATM-0716512, ATM-0745744, and AGS-0847126 grants. The National Center for Atmospheric Research is sponsored by the National Science Foundation. The sunspot models utilized in this investigation were computed at the National Institute for Computational Sciences (NICS) under grant TG-AST 100005.

**References**
Abramenko, V., Yurchyshyn, V., Goode, P., & Kilcik, A. 2010, ApJ, 725, L101
Berdyugina, S. V., Solanki, S. K., & Frutiger, C. 2003, A&A, 412, 513
Bharti, L. 2007, ApJ, 665, 79
Bharti, L., Beck, B., & Schüssler, M. 2010, A&A, 510, A12
Cao, W., Gorceix, N., Coulter, R., Ahn, K., Rimmele, T. R., and Goode, P. R. 2010, Astron. Nachr. 331, 636
Crocker, J.C. & Hoffman, B.D. 2007, Methods in Cell Biology, 83, 141
Crocker, J. C., and Grier, D. G. 1996, J. Colloid Interface Sci. 179, 298
Danielson, R. 1964, ApJ, 139, 45
Ewell, M.V. 1992, SolPhys, 137, 215
Grossmann-Doerth, U., Schmidt, W., & Schroter, E. H. 1986, A&A, 156, 347
Goode, P. R., Yurchyshyn, V., Cao, W., Abramenko, V., Andic, A., Ahn, K., & Chae, J. 2010, ApJ, 714, L31
Kitai, R. 1986, SolPhys, 104, 287
Kitai, R., Watanabe, H., Nakamura, T., Otsuji, K., Matsumoto, T., Ueno, S., et al. 2007, Publ. Astron. Soc. Japan 59, 585
Leenaarts, J., Rutten, R. J., Sütterlin, P., Carlsson, M., & Uitenbroek, H. 2006, A&A, 449, 1209
Muller, R. 1973, SolPhys, 29, 55
Riethmüller, T.L., Solanki, S.K., Zakharov, V., & Gandorfer, A. 2008, A&A, 492, 233
Rimmele, T. 2004, ApJ. 604, 906
Rimmele, T. 2008, ApJ. 672, 684
Schlichenmaier, R. 2009, SSRv, 144, 213
Schüssler, M. & Vögler, A. 2006, ApJ, 641, L73
Rempel, M., Schüssler, M., Cameron, R. H., & Knölker, M. 2009a, Science 325, 171
Rempel, M., Schüssler, M., & Knölker, M. 2009b, ApJ 691, 640
Rempel, M. 2011, Proceedings IAU Symp. 273, 8
Sobotka, M., Brandt, P.N., Simon, G.W.: 1997, A&A, 328, 689
Sobotka, M., & Hanslmeier, A. 2005, A&A, 442, 323
Sobotka, M. & Jurcak, J. 2009, ApJ, 694, 1080

Sobotka, M., & Puschmann, K. G. 2009, A&A, 504, 575
Socas-Navarro, H., Martinez Pillet, V., Sobotka, M., & Vazquez, M. 2004ApJ, 614, 448
Solanki, S.K. 2003, A&ARv, 11, 153
Tritschler, A. & Schmidt, W. 2002, A&A, 388, 1048
Vernazza, J. E., Avrett, E. H., & Loeser, R. 1981, ApJS, 45, 635
Watanabe, H., Kitai, R., & Ichimoto, K. 2009, ApJ., 702, 1048
Watanabe, H., Tritschler A., Kitai, R., & Ichimoto, K. 2010, SolPhys., 266, 5
Wiehr, E., & Degenhardt, D. 1993, A&A, 278, 584
Wöger, F., & von der Lühe, O. 2007, Appl. Opt., 46, 8015